\documentclass[11pt]{article}          % twocolumn, pdftex
\usepackage[paper=a4paper,left=25mm,right=25mm,top=25mm,bottom=25mm]{geometry}

\RequirePackage[T1]{fontenc}

\RequirePackage{graphicx}
\RequirePackage{mathptmx}      % use Times fonts if available on your TeX system
\RequirePackage{flushend}
\RequirePackage[numbers,sort&compress]{natbib}
\RequirePackage[colorlinks,citecolor=blue,urlcolor=blue,linkcolor=blue]{hyperref}

\usepackage{mathptmx}       % selects Times Roman as basic font
\usepackage{helvet}         % selects Helvetica as sans-serif font
\usepackage{courier}        % selects Courier as typewriter font

\usepackage{type1cm}        % activate if the above 3 fonts are
\usepackage{authblk}

\usepackage[bottom]{footmisc}% places footnotes at page bottom

\usepackage{subcaption}
\usepackage{svn-multi}

\usepackage{savefnmark}

\usepackage{comment}

\makeindex             % used for the subject index
\begin{document}
\title{Online Production Validation in a HEP environment}

\author{T. Harenberg}
\author{N. Lang}
\author{P. M\"attig}
\author{M. Sandhoff}
\author{F. Volkmer}
\affil{Bergische Universit\"at, Gau\ss{}str. 20, 42119 Wuppertal, Germany}  
\author{T. Kuhl}
\affil{DESY Zeuthen, Platanenallee 6, 15738 Zeuthen, Germany}
\author{C. Schwanenberger}
\affil{DESY Hamburg, Notkestr. 85, 22607 Hamburg, Germany}

% Use \authorrunning{Short Title} for an abbreviated version of
% your contribution title if the original one is too long
%\institute{Bergische Universit\"at Wuppertal, %\email{peter.mattig@cern.ch}}
%
% Use the package "url.sty" to avoid
% problems with special characters
% used in your e-mail or web address
%

\renewcommand\Affilfont{\itshape\small}
%\date{Received: date / Accepted: date}

%\noindent author: \svnauthor

%\noindent svn revision: \svnrev

%\noindent svn date: \svndate
\maketitle

\abstract{
Petabytes of data are to be processed and stored requiring millions of CPU-years  
in high energy particle (HEP) physics event simulation. This enormous demand
is handled in worldwide distributed computing centers as part of the 
LHC computing grid. These significant resources require a high quality 
and efficient production and the early detection of potential errors.
In this article we present novel monitoring techniques in a Grid environment to collect
quality measures during job execution. This allows online assessment
of data quality information to avoid
configuration errors or inappropriate settings of simulation parameters 
and therefore is able to save time and resources.}

\section{Introduction}
\begin{figure*}[t]
\begin{center}
\includegraphics[width=0.9\textwidth]{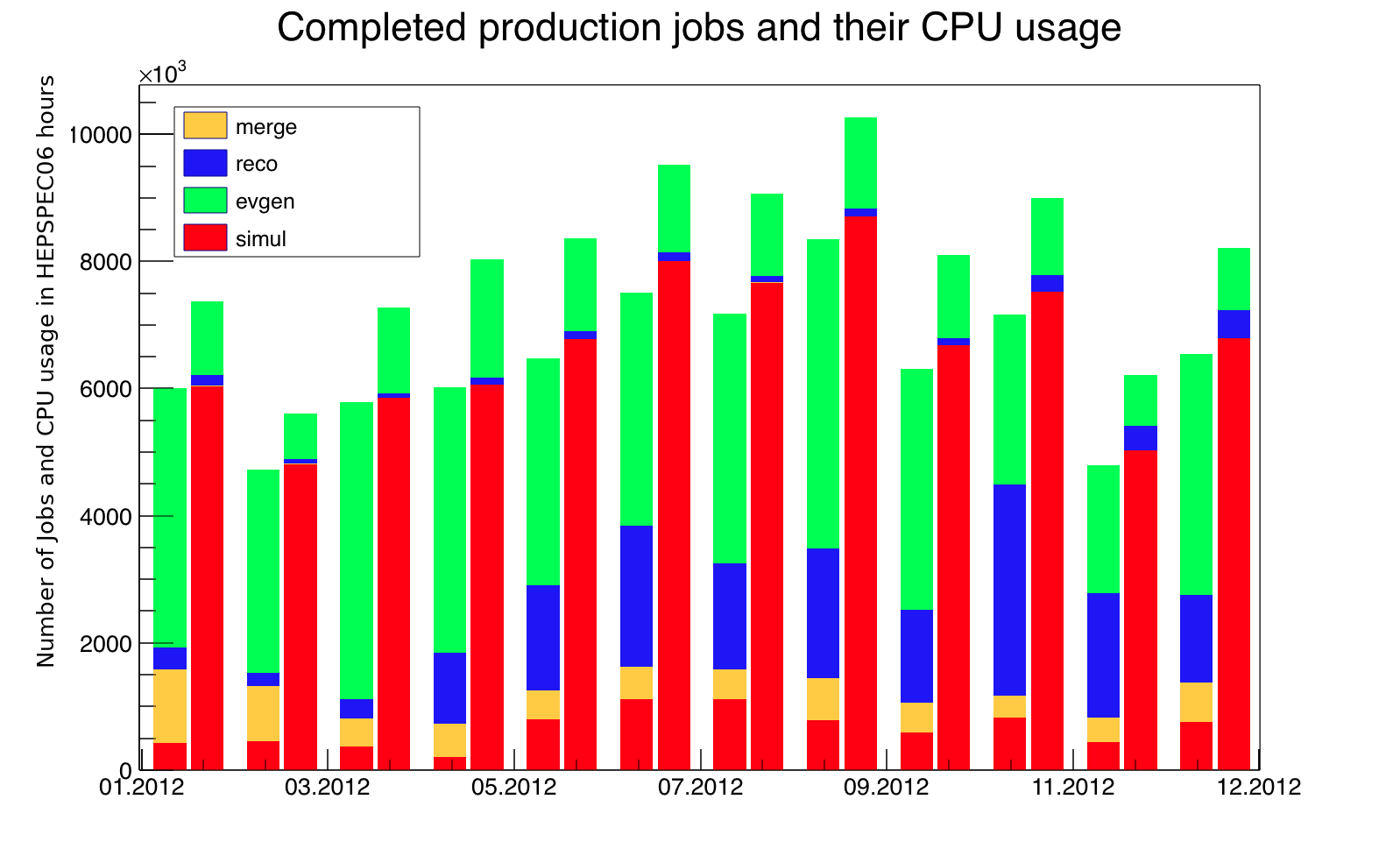}
	\caption{Number of ATLAS jobs (left bar) % (in millions) 
	per month 
	and their CPU consumption (right bar) for 2012. 
	The largest fraction of both  the number of jobs
        and the overall CPU consumption is due to detector simulation jobs (simul). Also shown
        is the corresponding usage for event
        generation, merge and reco type jobs (Numbers derived from~\cite{Karavakis:2014aia}).\label{img:resource-usage}}
\end{center}
\end{figure*}

Today's particle physics experiments are producing a huge amount of data 
both from direct measurement and simulations. The latter is 
mandatory to compare theoretical expectations to measured
data~\cite{Aad:2010ah}
and requires a detailed description of both the underlying physics processes and
the particle interactions in the detector. This process demands
significant amounts of computing resources~\cite{Karavakis:2014aia}, which
are provided by LHC computing centers distributed worldwide, the basis of the LHC
computing Grid.
As an example, in 2012, the ATLAS experiment 
at the Large Hadron Collider (LHC) has
simulated about ten trillion events\footnote{Each event represents the detector information on all particles
produced in a single particle collision.} stored in more than 54 PB of disk space,
and more than 100 000 jobs are processed every day. 
Figure \ref{img:resource-usage} shows a monthly distribution of computing
jobs\footnote{A job is one collection of programs processing the same
  physics process. 
%  It is tuned to run in a batch system friendly time
%  frame (typically 8 hours). 
%Collections of jobs are organized in tasks.
}.  Any error in 
processing this vast amount of simulation wastes CPU time and should be avoided.

The requirement of high efficiency and reliability in the production of these simulated events
is somewhat at odds with the very high complexity of the simulation programs, which therefore 
contain a high number of potential sources of mistakes.
The simulation programs are configured by a large number of settings 
describing special physics processes, special particle decay modes, theoretical models  
and the state of the detector at various energy levels and conditions.

Inappropriate settings may lead to job aborts, which would be easy to detect.
However, more difficult to identify are those faulty settings that cause the jobs 
to terminate technically without errors,  
but lead to results, which may be useless or even misleading for
physics analysis. The complexity and large range of running conditions 
also imply that inappropriate settings
are frequently discovered only after large samples have been produced and intensively used 
in analyses, wasting a substantial amount of computing and human resources and slowing 
down results. This requires a rapid employment of data quality tools and a prompt
and as much as possible automated response to potential failures. 

We argue in this paper that problems can be detected promptly during job processing 
by an on-line monitoring of production jobs within the grid environment and discuss a tool
for this: the 'Job Execution Monitor' (JEM)~\cite{diss:dossantos, dossantos:2012}. 
As a side effect, using a tool like JEM leads to a job-based resource monitoring. 

In this paper we present the application of such a monitoring program
which is used, for example, in ATLAS simulation.

% ==========
\section{Replacing manual by automized Monitoring}

\begin{figure*}[t]
\begin{center}
\includegraphics[width=0.9\textwidth]{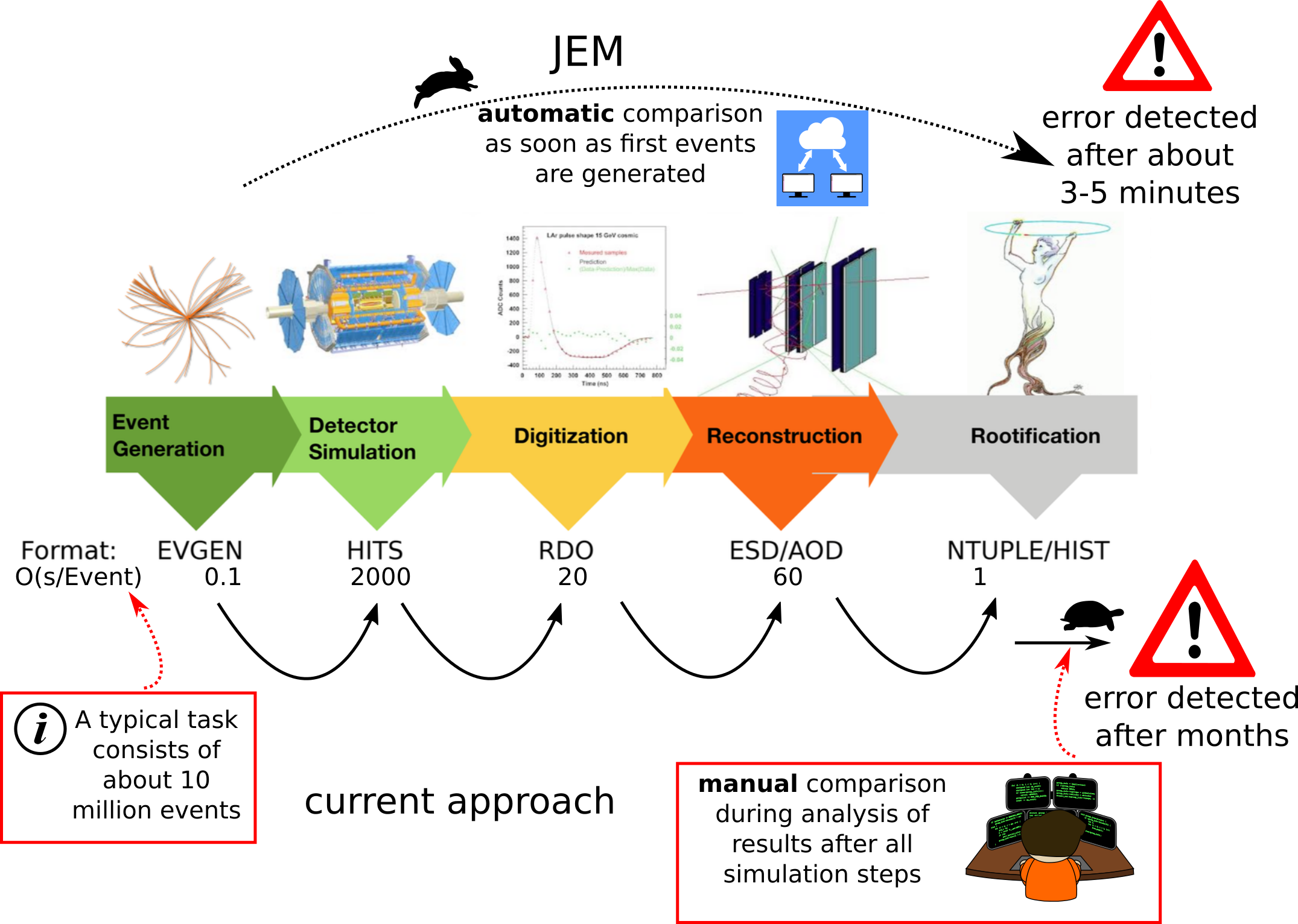}
\caption{
%Comparison of validating simulated events: left standard procedure, right new automated method. 
%Here a potential error in the event generation stage is assumed.
%Also shown is the typical computing times per event required for the various steps and the
%total time assuming a typical $10^7$ events to be generated per task. 
%Note that the required time for the analysis step depends on the kind
%of analysis 
Schematics of the steps of a simulation task indicating the advantages of the 
validation strategy discussed in this paper: while up to now all simulation steps have to be
passed and a possible problem can only be detected by chance in a
final analysis, this work proposes an automated validation which is
started during the first simulation step
(graphic partly derived from~\cite{Debenedetti:1605835}).}
\label{img:simu-valid}
\end{center}
\end{figure*}

% TK: check citations
The major steps in physics simulation~\cite{Aad:2010ah} at the LHC
towards its use in analysis are depicted in
Fig. \ref{img:simu-valid}.
In this figure the sequence of processing steps is listed. 
In the first two step events are generated using theoretical models,
and particles from this generator stage are tracked in the detector,
repectively decays and interactions in the
detector material are simulated.
In a third step they are digitized, i.e the response of sensors to the 
interacting particles is modeled. All these steps use Monte Carlo (MC) techniques.
Finally physics objects are reconstructed
from these signals and the results converted into a physics analysis data format.

To obtain sufficient statistics as well as to check for rare event topologies, 
tens of millions of events of a certain generator 
configurations have to be produced. The production of events with identical
configurations will be denoted as 'tasks'. 
To minimize resource losses due to technical failure, 
each task is parallelized into many computing jobs, 
typically of 5000 events each.

The tasks are processed serially, i.e. each step has to be finished for all of the generated events, 
before the next step is started. 
This leads to a significant latency between the first event generation and its ultimate use 
in physics analysis can be performed.
It also means that a significant amount of resources is employed.

As a consequence, problems, e.g.~those introduced in the first step by inadequate 
parameter settings, are only identified after all events have been processed. 
For the most part experiments rely on the effort of individual members to check manually many
histograms in order to verify the integrity of the results.
%While being helpful, it falls short of solving the problem in general 
%due to various reasons.
To perform such evaluation reliably 
requires a substantial effort, it is time consuming and requires a 
high level of expertise.
Furthermore, comprehensive checks would at best require a significant amount of time
and resources. However, this is most often prohibitive given 
the amount of information that needs to be digested.
% Effectively validation results are only available after the whole
% simulation chain has been completed. 
Therefore this monitoring by hand is delayed and susceptible to failure.
In fact, it may happen that such problems are detected only after the simulation results
have been used for several months in the analysis. 

Instead of manual and lengthy tests, the method that is discussed in this paper is 
based on two main ingredients:

\begin{itemize}
\item the checks are automated with a central tool, 
\item they are performed during job execution on the grid.
\end{itemize}

\noindent Using such a method provides immediate response
and allows one to monitor all kind of relevant data quality information.
Moreover, the criteria to identify problems are unambiguously defined and agreed upon.
Therefore the proposed automated procedure
only needs to be developed once and can be centrally maintained.
Monitoring of production jobs can thus be significantly simplified and accelerated to secure 
an almost immediate response. Furthermore, the observables to be monitored can be expanded
without major cost and effort.

In the framework introduced here, these general data quality potentials are
applicable on - line and in a grid environment.  
The procedure can therefore  
be used to rapidly trigger a termination of jobs, if significant problems are identified 
thus preventing a waste of computing resources. 
The procedure is schematically shown for the case of monitoring the generator 
stage in Fig. \ref{img:simu-valid}, where early monitoring has the highest benefit, as errors can be detected before the costly detector simulation stage.

This new concept can be realized in JEM,
which has been developed for online and remote data monitoring. While JEM 
has a wide range of features, in the context of this application,
it uses a set of reference histograms and compares them to their respective counterparts
while these are being generated in the job.

% ==============
\section{JEM}

Before discussing its application to generator validation, the general concept of JEM will be 
summarized.
JEM was originally developed for online monitoring of jobs in a grid environment, where it aimed
at identifying causes of execution problems.

The Job Execution Monitor has features to help identifying the reasons for job failures or irregular job 
behavior based on different metrics taken during job runtime.
These metrics include consumption of CPU, memory and disk space, the number of open 
file descriptors and network throughput.
Advanced analysis techniques like online log file analysis and stack trace inspection are   
available, together with the ability to follow step by step execution of Python and Bash scripts and of binary files that have been prepared before submitting the job to the grid.

JEM is mainly developed using the Python interpreter\footnote{The Python part has grown
to almost 230.000 lines of code.}, with few core components  
written in the programming language {\tt C}.
The web services and database backends are based on the Django
framework~\cite{url:django}. 

In addition to the functionality of the job instrumentation, JEM
provides several services at a central machine called "JEMserver".
These services include data management, like a database for job monitoring data
storage, a web server to display 
monitoring and validation results and a central file cache.

% ========
\section{Applying JEM to validation of production}
\label{sec:application}

The large potential of JEM can be applied to online data quality.
The validation procedure with JEM is based on the comparison of 
histograms and it has three components. The first one are certified histograms from
previous tasks, denoted 'reference histograms',  
which are relevant for the task that is to be validated 
and stored in the central file cache. 
The second
component are histograms which are produced during the processing of the new
task and are denoted 'quality histograms'. The production of these histograms and
the statistical comparison of reference and quality histograms is performed 
automatically. In case of a process considered to be
identical to a previous reference, an immediate evaluation is provided. In case of the
more frequent case that the parameters in the production have been changed even
slightly, as a third component a physicist has to ultimately judge 
the severity of possible disagreements. These steps and the detailed integration into JEM
will be discussed next.

% =========
\subsection{Categories of histograms}
\label{sec:histocategories}

\begin{figure*}[t]
\begin{center}
\includegraphics[width=0.9\textwidth]{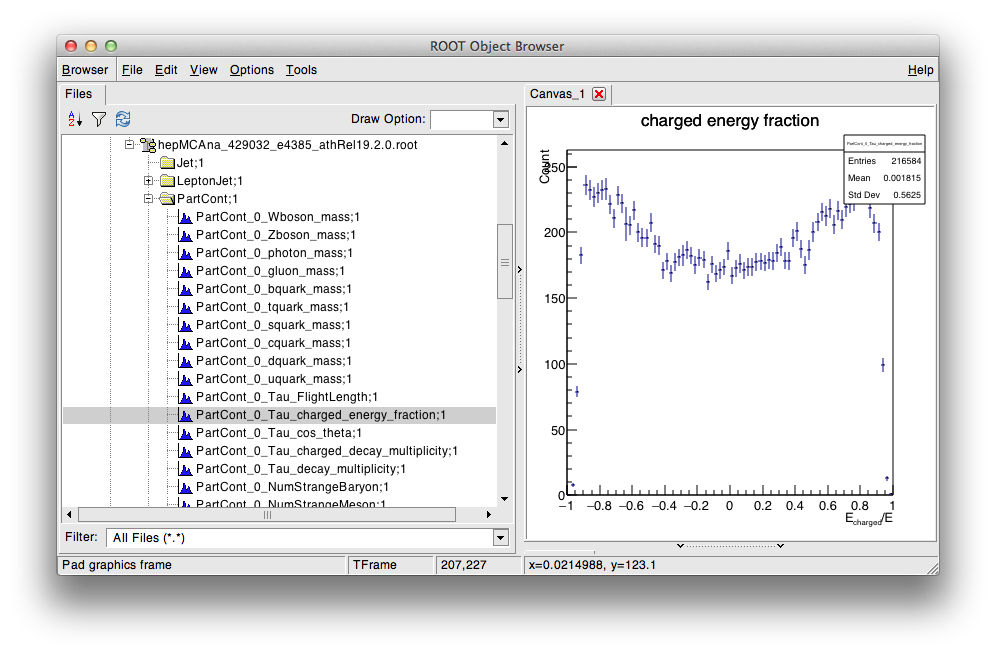}
\caption{ROOT browser showing parts of the topological structure of a HepMCAnalysis output file.} 
\label{img:tfile-example}
\end{center}
\end{figure*}

A key issue for the comparison of the simulation tasks is the definition of which distributions to 
compare. Two standardized production frameworks for histogramming at the generator level
are invoked up to now: HepMCAnalysis~\cite{hepmcana} and
RIVET~\cite{RIVET}\footnote{RIVET is a widely used analysis package 
in particle physics that contains
comparisons of high energy physics data with MC simulations for a
large number of processes and analyses. 
% These analyses are available now
%within JEM for the purpose of comparing MC reference files with MC
% validation files. Since RIVET is  in the field, the
% interface to JEM represented an important development that was a high
%priority for MC validation at ATLAS.
}
Both produce ROOT~\cite{root} histogram collections organized in
subdirectories for differently defined analysis types. An example is
shown in Fig. \ref{img:tfile-example} for HepMCAnalysis. 
% Rivet and HepMCAnalysis are the first analysis codes fully implemented
% in JEM. 
Expanding JEM validation to more analysis packages can easily be achieved by
simply providing a small glue code.

% ==
% ===

% =========
\subsection{Relevant JEM infrastructure}

\begin{figure*}[t]
\begin{center}
\includegraphics[width=0.9\textwidth]{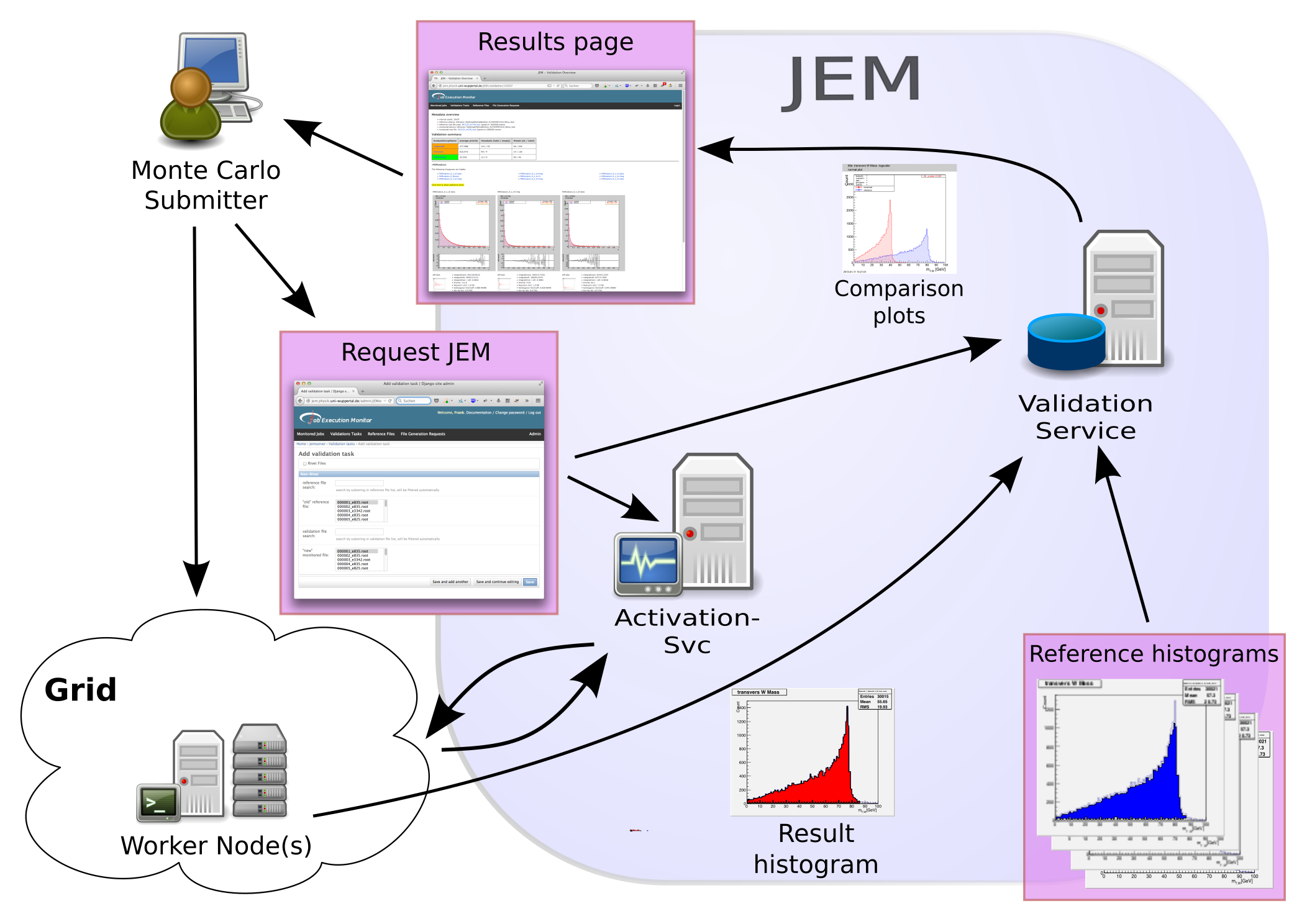}
\caption{Schematic of JEM's job validation in the grid. A
  Monte Carlo submitter sends a task into the grid and has to enter the task identifying data into the Request JEM web interface. From this interface, the Activation Service and the Validation Service are configured   with the number of jobs to be instrumented and the choice of reference histograms to be used. 
Instrumented jobs produce quality histograms and send these to the Validation Service, where they are compared
against the reference set. The comparison result can then be accessed via a web page.} 
\label{img:jem-validation}
\end{center}
\end{figure*}

The JEM Activation Service is the kernel of the infrastructure (see fig.\ref{img:jem-validation}).
It runs as a central service on the JEMserver, is contacted by every
grid job, which is about to start, and decides whether this particular job should
be instrumented with JEM or not. 
In the current test version this decision is steered manually by a web interface, and
the monitoring jobs run just after the production task is finished.
This service has been optimized for speed, database access is cached and the amount of exchanged data is very small. It could be proven that it holds the load of about 11 jobs per second. Furthermore, a fallback procedure in every job prevents it from failing even if the central server would not respond within a reasonable time frame. 
Typically within each task around 200 jobs 
will be instrumented to obtain about a million events allowing
sufficient statistical power for each histogram.
The additional time overhead needed to run the instrumented jobs is typically of
the order of minutes and therefore negligible 
compared to the total runtime of several hours needed for the
simulation and reconstruction of events.
A rule engine decides which job will be instrumented, based on the
given meta data of the job that is accompanying each request.  
This meta data  includes, among others, information about the location of the job,  
the submitter, the job type and the job identifier. 
%A high granularity selection can therefore be realized, combining detailed information from the jobs' meta data. 

As the JEM monitoring component is acting as a payload job's wrapper,
this has full control over its child  processes. 
Once it detects a successfully finished job, further actions can be
initiated without notifying the outer batch system about the job's end,
which is especially important for the validation, as will be described
in Sec.~\ref{ssec:jem-validation}.

The schematic workflow is 
shown in Figure \ref{img:jem-validation}, and it is explained next.

% ======
\subsection{Initiating JEM Validation}

The  validation process within the JEM system is initiated using  
the JEMserver web interface\footnote{\url{http://jem.physik.uni-wuppertal.de/}}.
This initialization step requires an authorized person with a valid account that guarantees data protection measurements by the experiments.
The required input data includes the task identifier, the desired number of jobs
to be instrumented, and the keywords to select the respective physics
analysis module. 

The inputs are stored in a database and several services on the
JEMserver then either add appropriate rules to the JEM Activation Service or store a 
search pattern that is regularly executed to find a task matching the given criteria.
The rules in the Activation Service keep track of the number of jobs that have been 
instrumented.

%The JEM Validation Service is also configured with the reference histograms and set to 
%expect incoming monitoring data in the future.

% ====
\subsection{Quality Histograms}
\label{sec:histo}

Once jobs have been instrumented with JEM the quality histograms of
the same type and format as 
the reference histograms are created. This comprises more than 200
histograms.
%As the output files from the just finished job are still available in
%the local working directory, these files are used to
%create the quality histograms with a small analysis script launched by JEM. 
The production of quality histograms can be performed online during
job execution
or offline after the specified jobs have terminated.
% The runtime of theses analysis scripts are usually quite small
% (ca. 3 to 4 minutes), which is a negligible time overhead compared
% to the job runtime, which typically takes several hours. 

% =========
\begin{table*}[t]
\centering
\begin{tabular}{p{0.2\textwidth} p{0.4\textwidth}}
\hline \hline \\[-2ex]
\textbf{field} & \textbf{description} \\[0.5ex] \hline \\[-1.8ex]
fileName & physical file name on disk \\
dataSetId & ATLAS specific; describing the underlying physics process \\
eTag & software tag: ATLAS specific; a tag associated with a software cache, containing specific software versions \\
swGenTag & version of histogram generation software \\
swConfig & configuration parameters for  histogram generation software \\
physRef & description of the underlying physics process  \\
ts & timestamp of creation or last update \\
eventCount & number of events the file is based upon \\
generatedWith & histogram generation software \\

\\[-1.8ex] \hline \hline
\end{tabular}
\caption{Metadata as stored in the database per file, both for
  reference and quality histogram files.
%\PMnote{\bf What exactly do these metadata correspond to? Ref histo? Qual hist? General?}
}
\label{tab:reffile-metadata}
\end{table*}
% =======

Once the quality histograms have been created, they are sent to the JEMserver where  
they are merged and stored in the validation file cache. A database entry is added
for each quality histogram collection file  
(an excerpt of its content is listed in Table \ref{tab:reffile-metadata}).
This file is also stored at the JEMserver to serve as reference file
for future validations. 

\begin{comment}

\begin{table}[t]
\centering
\caption{Example database entries of each merged histogram file that
  can be specified in the web interface for validation.}
% CS: letzte Spalte loeschen!
\label{tab:reffile-metadata}
\begin{tabular}{p{0.2\textwidth} p{0.4\textwidth} p{0.4\textwidth}}
\hline \hline \\[-2ex]
\textbf{field} & \textbf{description} & \textbf{example} \\[0.5ex] \hline \\[-1.8ex]
fileName & physical file name on disk & hepMCAna\_\-117581\_\-e3841\_athRel19.2.0.root \\
dataSetId & ATLAS specific; describing the underlying physics process & 117581 \\
eTag & software tag: ATLAS specific; a tag associated with a software cache, containing specific software versions  & 3841 \\
swGenTag & version of histogram generation software & 19.2.0 \\
swConfig & configuration parameters for  histogram generation software & \{"JetAnalysis": true, "ParticleContentAnalysis": true, "LeptonJetAnalysis": true, "PdfAnalysis": true\} \\
physRef & description of the underlying physics process & Sherpa\_\-CT10\_\-ttbar\_\-dilepton\_\-MEPS\_\-NLO \\
ts & timestamp of creation or last update & Dec. 7, 2015, 10:55 a.m. \\
eventCount & number of events the file is based upon & 8.000.000 \\
generatedWith & histogram generation software & HA \\

\\[-1.8ex] \hline \hline
\end{tabular}
\end{table}

\end{comment}

% =======
\subsection{Validation process}
\label{ssec:jem-validation}
The comparison between the monitored and the reference histograms is performed
by the JEM Validation Service, 
which is a worker process running on the JEMserver that reacts to state 
changes of validation tasks in the database. 
For visualisation purposes and automatic quality control, 
a standard data quality program, DCube~\cite{url:dcube}
%\footnote{DCube is a ROOT~\cite{root} based
% software for histogram comparisons.} 
is launched to perform a
statistical analysis to estimate the level of agreement between
monitored and reference file. This agreement is calculated using
one or more statistical tests. Currently these are
the Kolmogorov-Smirnoff test and Pearson's $\chi^2$.
Each of these statistical tests $t$ returns a p-value $p_t$. %in the interval $[0:1]$.
% defining the level of agreement of the two histograms.
An overall estimator, the comparison severity $S$, is calculated using the weighted 
sum of the individual test results for each histogram comparison,

\begin{equation}
       S\  = \frac{\sum_{\mathrm{t}} (1 - p_t) \cdot w_{\mathrm{t}}}{\sum_{\mathrm{t}} w_{\mathrm{t}}} \,\, ,
\label{eq:severity}
\end{equation}
\noindent where %$R_t$ is the value of the statistical measurement $t$, and
$w_{\mathrm{t}}$ is a weight factor that can be set to give preference to certain test results,
and whose current default values are set to 1 for both tests.
%given in Table~\ref{tab:dcube-weights}.
$S$ is a simple measure of the inconsistency between both histograms.
Histograms are considered consistent and therefore 'ok', if $S \ <
0.25$. Values of $0.25 < S < 0.5$ would be considered as warning, and
values above 0.5 are problematic and should be reviewed.  
It further allows one to present all histograms of an analysis group in an ordered fashion, 
e.g. having those with the largest $S$, i.e. the most inconsistent histograms first.

% =======
\subsection{Validation output web page}

\begin{figure*}[t]
\begin{center}
\includegraphics[width=0.8\textwidth]{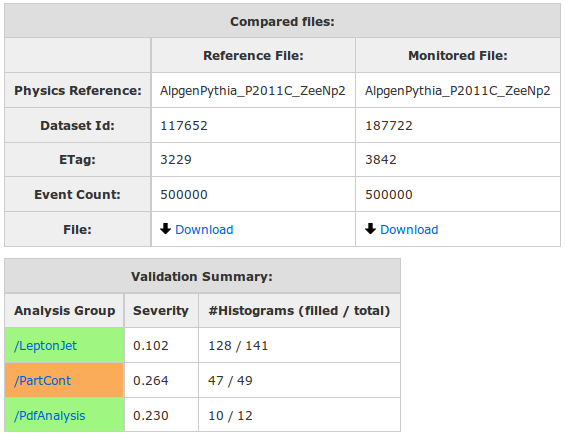}
\caption{
Top of a result web page. An overview of the metadata is given as well as 
a summary of the validation result. The left column of the lower table gives
the name of the physics analysis, the centre column lists the mean
of the severity values of all histograms calculated using 
Eq.~\ref{eq:severity}, leading to the colour coding for each analysis group.
The right column presents the number of filled and the
number of total histograms produced. Histograms can be empty depending
on the information accessible in the MC generator file and on
the physics process investigated. Empty histograms are ignored for the
calculation of the mean severity.} 
\label{fig:result-hp1}
\end{center}
\end{figure*}

The results are stored in a file, containing all results on the comparisons. 
In order to easily interpret results,
they are graphically presented on a web page, an example of which is shown in 
Fig.~\ref{fig:result-hp1}. 
An overview of the metadata is given at the top. 
A summary of the quality of the validation is presented in classes
according to the analysis groups. The information contains
the average $S$ value ('severity') and the number of non-empty and total
histograms.
The global agreement can be easily assessed since the outcome is 
presented using a color code of green, orange and red
depending on the number of failed tests.
% These features came out to be of tremendous help to find potential problems in
% the investigated MC samples. 

% ======
% ========

\begin{figure*}[t]
\begin{center}
\includegraphics[width=0.9\textwidth]{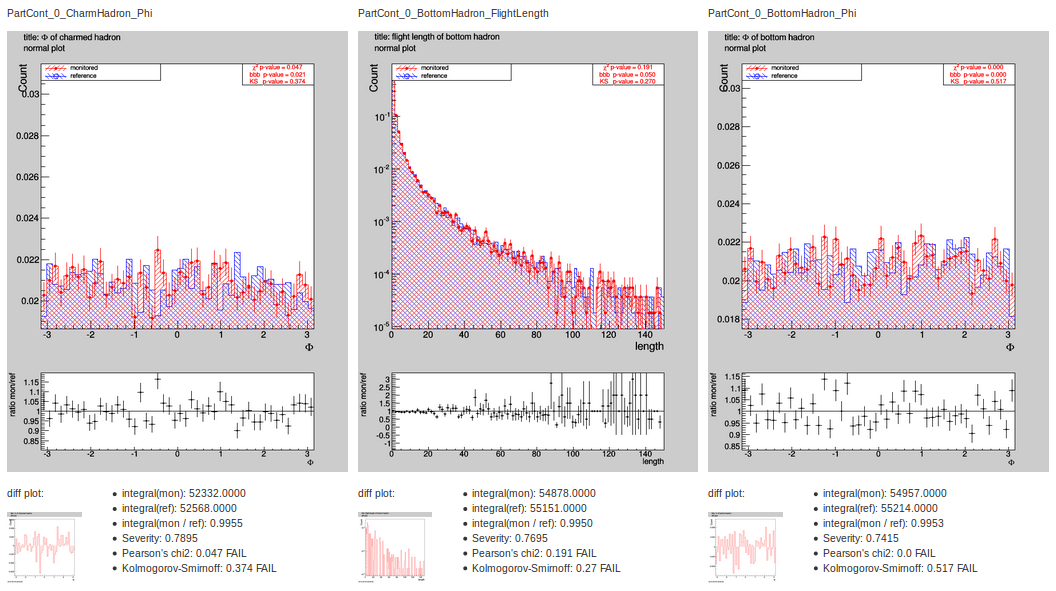}
\caption{
Result web page of a specific analysis module containing the
  plots ordered by the severity $S$ 
of the disagreement between quality and reference histograms. 
The ratio of the quality and the reference histograms is shown beneath each
histogram. Also a small version of the difference plot is shown,
which can be expanded to show the absolute difference between the two
histograms. For each plot the statistical agreement
is listed, together with the normalization scale.
The three plots with the highest severity are shown together with
their statistical test results. All other plots of an analysis group
are hidden, but can be expanded on demand by clicking on respective links.  
} 
\label{fig:result-hp2}
\end{center}
\end{figure*}

A more detailed information of potential disagreements is
provided by showing the individual distributions. 
The histograms for the specific analysis modules are also listed (see
Fig.~\ref{fig:result-hp2}). Within each module they are ordered
according to the magnitude of the severity $S$ of the disagreement. 
Therefore the histograms with the largest discrepancies are
displayed on the top. This ensures that the most problematic histograms can be 
spotted easily. The level of (dis)agreement can be assessed since 
each quality histogram is overlayed to its reference histogram. 
Also shown is the ratio between these histograms. 
Furthermore, the results of the statistical tests are listed.

\begin{comment}
\begin{table}[t]
\centering
\caption{Histogram comparison analysis weights.}
\label{tab:dcube-weights}
\begin{tabular}{p{0.33\textwidth} p{0.2\textwidth}}
\hline \hline \\[-2ex]
\textbf{Histogram comparison} & \textbf{weight factor} \\[0.5ex] \hline \\[-1.8ex]
Kolmogorov-Smirnoff & 900 \\
Pearson's chi2 & 90 \\
bin-by-bin & 9 \\
Mean Test & 0.99 \\
\\[-1.8ex] \hline \hline
\end{tabular}
\end{table}
\end{comment}

% % =====
% \subsection{Presentation of results and final assessment}

% The results are stored in a file, containing all comparison results. To interpret them easily,
% they are graphically presented in a web page, an example is shown in Figure \ref{fig:result-hp2}.
% This page contains both
% a summary of the fraction of histograms with major problems as well as those deemed ok,
% and a detailed display of all quality histograms and comparing them to the reference histogram.
 
% To easily assess the level of agreement, the basic outcome is presented in a color code 
% representing potential problems or agreement.
% The results are classified according to the analysis groups. 
% An overview over the relevant files is also provided, in particular
%links to the sets of quality and reference histogram files for more detailed inspection.

The fast turnaround and the immediate availability of the results on a web page 
provide an easy and fast tool for validation purposes.
All necessary actions run automatically in the background and can also be easily 
configured using a web interface.

Although these validation steps run automatically, the assessment from experts is required 
in order to judge 
the importance of potential differences. This manual judgement is needed since very rarely
tasks are submitted under identical conditions, but  differences are expected due to
changes in parameter settings or new versions of generators to be validated.

\section{Operational experience}

In the two years since the first prototype of the JEM Validation Service was launched manually,
more than 540 different task validations 
have been initiated by the users of the ATLAS production system.
This represents 197,660 individual histogram comparisons that had 
to be evaluated.
Overall the JEM Validation Service created 409 quality histogram files 
based on 345,703,764 events. 
As described in Sec.~\ref{sec:histocategories}, two different publicly
available analysis software tools,
HepMCAnalysis~\cite{hepmcana} 
and RIVET~\cite{RIVET},
are used to create the histograms to compare a reference MC sample
with the monitor MC sample to be validated.
In both cases, JEM is very simple to use, therefore the
ATLAS Collaboration employed offline shift personnel to overview the
validation, which was possible with only a small amount of training.

% Different analysis software was used to create the histograms that
% were used by JEM to compare a reference sample with the sample to be
% validated. Different analysis modules appropriate for different physics
% processes can be selected. 

In practice the very large majority of the tasks showed no problematic distributions. 
The ease to interpret results, e.g. the colour coding,  
has saved the MC validation shifters a significant amount of time and resources. This made 
JEM crucial for a fast and reliable use of simulation. In addition some tasks
led to inappropriate results and were quickly discovered using JEM. 
Some examples are given in Ref~\cite{ATLAS-PUB}. As will be discussed next,  
JEM helped to quickly validate new versions of Monte Carlo generators.

% ======
\subsection{PYTHIA validation}

\begin{figure*}[t]
	\centering
	\begin{subfigure}[b]{0.35\textwidth}
		\includegraphics[width=\textwidth]{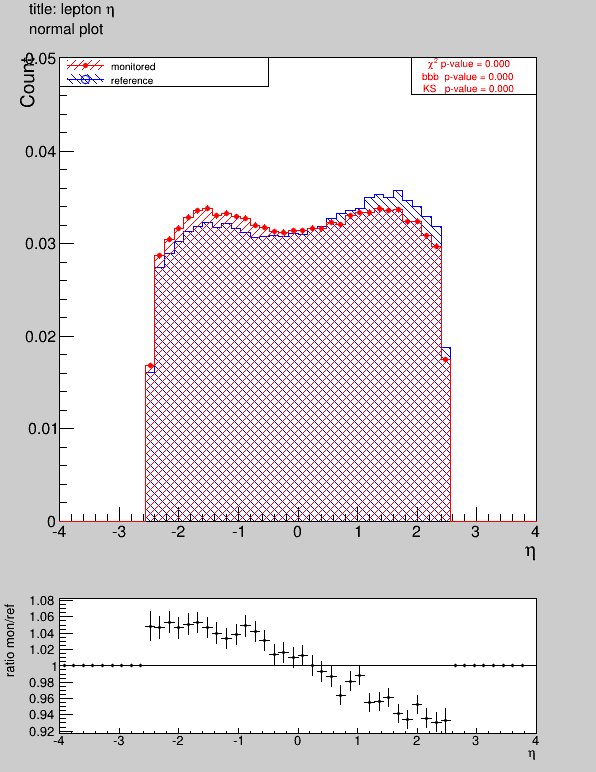}
	\end{subfigure}
	\begin{subfigure}[b]{0.35\textwidth}
		\includegraphics[width=\textwidth]{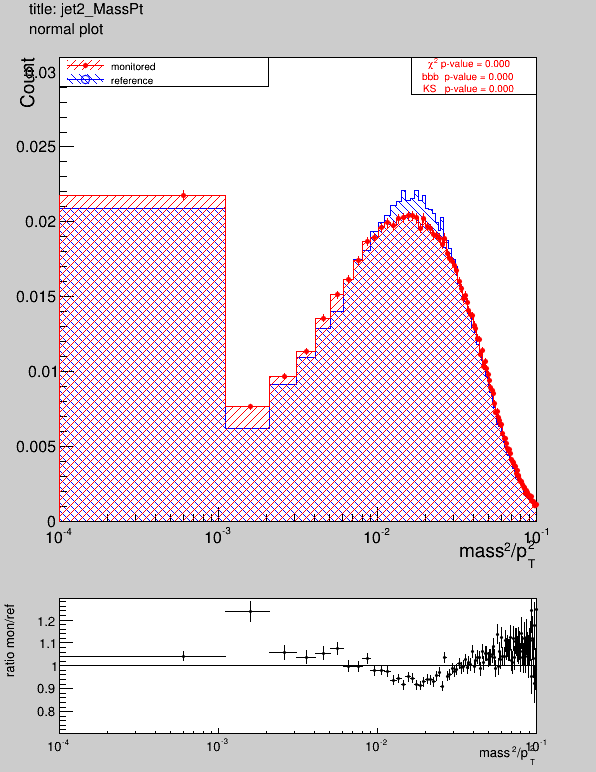}
	\end{subfigure}

	\caption{Example distributions
          comparing ALPGEN+PYTHIA~6.427 (blue) with
          ALPGEN+PYTHIA~6.428 (red) in simulated $Z \to ee$+2 partons events. Left: Pseudorapidity
          distribution of tight electrons; right: $M^2/P_T^2$ of the
          jets with second highest transverse momentum.
\label{img:failed-pythia-validation}}
\end{figure*}

One example involves the validation of PYTHIA~6.428~\cite{pythia}
which was compared 
to PYTHIA~6.427 as reference. This was the first validation campaign
of PYTHIA in the ATLAS collaboration using JEM. 
It became apparent
that the reference version PYTHIA~6.427 had problems that previously
remained undiscovered. This shows that the systematic validation with
JEM represents a large improvement compared to previous validation
methods. The blue distribution in
Fig.~\ref{img:failed-pythia-validation}, 
shows the the pseudorapidity distribution of the leptons that passed tight
selection requirements in the Drell--Yan production, $pp \
\rightarrow \ e^+e^- \ + \ 2~\mathrm{partons} \ + \ X$. 
The two additional partons in the matrix element were generated
using the ALPGEN generator~\cite{alpgen}, to match parton showering and 
matrix element partons the MLM procedure~\cite{MLM} was invoked
and the string fragmentation model was used to build primary hadrons from the final partons.

Apparently, the reference sample has a problem,
because the distribution is not symmetric, in contrast to the expectation.
The red distribution shows the same observable in the next PYTHIA version 
PYTHIA~6.428, showing that the problems were fixed in this version.

Only through the systematic investigation using JEM this
error was discovered by ATLAS. Consequently, the version PYTHIA~6.427 was
discarded from further usage. 

On the other hand, JEM was also used to quickly check if intended changes in the 
distributions are actually realised. An example from the two versions of PYTHIA  
is shown in Figure~\ref{img:failed-pythia-validation}
(right). Here the distributions of the jet mass squared $M^2$ 
over $p_T^2$ is displayed. This shows that both versions differ
in the simulation, a result that was expected since for the newer
version some changes in the PYTHIA string fragmentation simulation were implemented by
the authors. 

% =====
% ======

\subsection{SHERPA validation}

Another example where the systematic MC validation with JEM found previously unknown 
problems involved the validation campaign of {\sc  SHERPA}\cite{bib-SHERPA}. 

{\sc SHERPA}~1.4 was a multileg generator containing tree-level matrix
elements. For the $W$+jets production process this involved the 
multi-parton matrix elements for $W$+0~jets, $W$+1~jet, $W$+2~jets, and $W$+3~jets in
leading order (LO) Quantumchromodynamics (QCD) combined with the QCD
parton cascades following the CKKW merging scheme~\cite{CKKW}, and
fragmentation of partons into primary hadrons described using a
phenomenological cluster-hadronisation model. 

{\sc SHERPA}~2.1.0 was one of the first publicly available versions of this multileg
generator that included also matrix elements in next-to-leading order (NLO)
QCD. In case of $W$+jets production, this version contained matrix
elements of $W$+0~jets,
$W$+1~jet, and $W$+2~jets in NLO QCD, and $W$+3~jets, $W$+4~jets in LO
QCD, matched to parton showers using the CKKW 
method. This version represented a major step in developing MC
multileg generators including NLO QCD matrix elements. Since the
changes to previous versions were large, a very careful MC validation
was necessary. The systematic MC validation method using JEM was
essential in this task.

During the validation of {\sc SHERPA}~2.1.0 problems became obvious. Generating $W$+jets
events with subsequent $W\rightarrow e \nu$ decay in
proton--proton scattering at a centre-of-mass energy of 8~TeV, for example, displayed
asymmetric distributions of the pseudorapidity of hadrons including
charm or bottom flavour, while those distributions were expected to be symmetric as
predicted in {\sc SHERPA}~1.4. After private communication with the
{\sc SHERPA} authors, this problem could be traced 
to an error in the multiple parton interaction matrix elements, which was
fixed in version {\sc SHERPA}~2.1.1. 

\begin{figure*}[t]
	\centering
	\begin{subfigure}[b]{0.35\textwidth}
		\includegraphics[width=\textwidth]{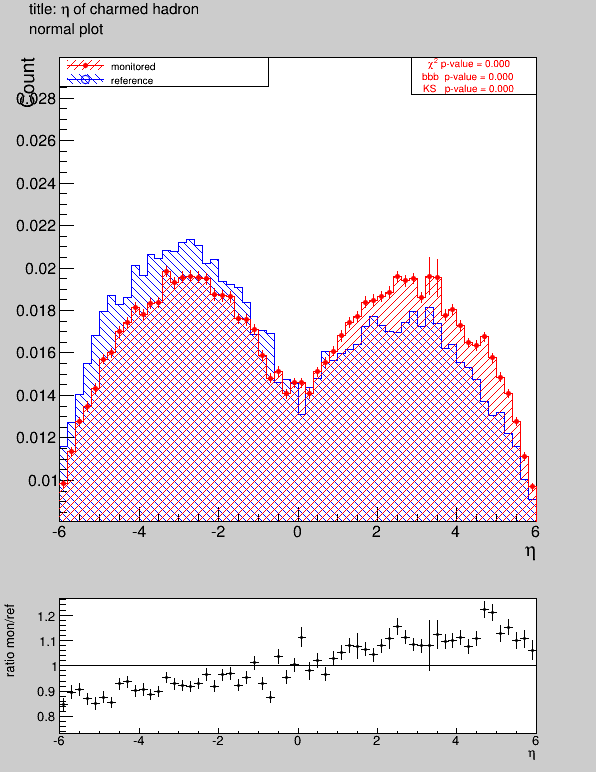}
	\end{subfigure}
	\begin{subfigure}[b]{0.35\textwidth}
		\includegraphics[width=\textwidth]{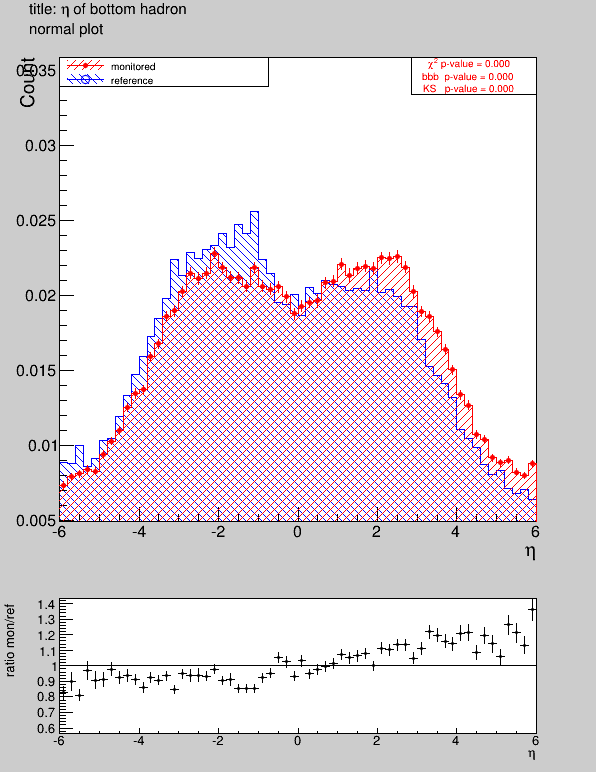}
	\end{subfigure}

	\caption[]{Distribution of pseudorapidity
          of hadrons with charm (left) and bottom (right) quark flavour comparing {\sc
            SHERPA}~2.1.0 (blue) with {\sc SHERPA}~2.1.1 (red)  for simulated 
          $W \to e \nu$+jets events. 
\label{img:failed-sherpa-validation}}
\end{figure*}

This
can be seen in Fig.~\ref{img:failed-sherpa-validation} where the pseudorapidity
distributions for charm hadrons (left) and for bottom
hadrons (right) are compared between the repaired version {\sc SHERPA} 2.1.1
(red) and the problematic version  {\sc SHERPA}~2.1.0 \\ (blue). As a result, in version {\sc SHERPA}~2.1.1, the pseudorapidity distributions of charm and bottom hadrons are
now symmetric.  
%The remaining differences compared to version {\sc
%SHERPA}~1.4.0 are due to the different matrix elements 
% as described above and due to 
%In a similar way a problem in the shower
%simulation of partons produced close to the beam axis was found which led to a modification
%in the next {\sc SHERPA}~2.2.0 version.

% ========
% =====

This is another example that through the systematic MC validation with
JEM, a prompt identification of issues within a new release of 
generators was possible. This information and the large set of
plots produced by JEM was of tremendous help for the SHERPA authors to
fix this and further problems.\footnote{In a similar way, for example, a problem in the shower
simulation of partons produced close to the beam axis was found which led to a modification
in the next {\sc SHERPA}~2.2.0 version.} Furthermore, it prevented mass production of 
mis-confugured samples by
the ATLAS and CMS collaborations and therefore saved an enormous amount
of CPU resources. 

\section{Summary \& Outlook}

In this paper a new validation tool based on the Job Execution Monitor (JEM) 
is presented. JEM is ready to run inside the Grid framework as a payload and automatically 
collects and compares quality measures of simulated physics processes during
event generation. Utilizing its current test version that allows manual steering
by a web interface, JEM has already become a standard to validate ATLAS simulation and 
gives ATLAS users the opportunity to detect inappropriate results due to 
mis - configurations in Monte - Carlo generators and unexpected features of simulation tasks.
It accomplishes this by automatically comparing and
classifying certified reference histograms with the new ones produced during
simulation. This aids in early discovery of problems and avoids a waste
of resources and time. It has also been extremely helpful in validating new versions
of QCD generators.

The results and the user interface are presented as a web service. With an
easy color code and sorted histograms, where only the most important ones are displayed, 
a quick assessment of the level of agreement can be
achieved. This mode of presentation was tested in practice and improved in collaboration
with the ATLAS Monte Carlo working group and their shifters. 
The overhead of generating the
validation plots and the automatic comparison of histograms is 
negligable compared to the runtime of typical simulation tasks.

% The paper shows that the early validation with JEM leads to a huge saving of CPU
% and working time, if errors are catched in the first simulation step
% and not after months of analysis work. 

% The JEM Validation Service has been established as a core component of
% ATLAS generator validation efforts.
% Special validation shifters are employed to initiate validation
% actions, inspect the results and communicate the feedback to the
% various generator development teams. 

It is currently investigated to automatically
instrument a small fraction of every generator task within ATLAS.
This would give quality histograms, which would be readily available
should the need arise to inspect them.

Whereas currently JEM is used at the early generator stage of the simulation chain,
its validation system can be expanded to other stages, such as simulation, digitization or
reconstruction and beyond. 
Also in this case a significant saving of time and resources
can be expected.
The simplicity of JEM installation and operation makes it
also possible to use it outside ATLAS and HEP experiments.

% Finally, it is planned to move the service to CERN's infrastructure.

\section{Acknowledgements}
We like to thank Leonid Serkin and Maria Moreno Llacer, together with the ATLAS generator validation team, for their invaluable feedback on using the JEM Validation Service.
It helped to improve the overall structure and user experience. 
We are grateful to Markus Elsing and Marcello Vogel for very helpful comments on
the text. Also, we would like to thank Frank Ellinghaus for discussions and
Ewelina Lobodzinska for the computing support and Nataliia Kondrashova
for the HepMCAnalysis modules.

\end{document}